\documentclass[a4paper, 12pt]{article}

\usepackage[]{babel}

\usepackage[
pdftex,
colorlinks,%
linkcolor=blue,citecolor=red,urlcolor=blue,
hyperindex,%
plainpages=false,%
bookmarksopen,%
bookmarksnumbered%
]{hyperref}
\usepackage{geometry}
\usepackage[backend=biber,style=numeric,sorting=none,doi=true,url=true]{biblatex}
\renewbibmacro*{editor+others}{%
  \ifnameundef{editor}
    {}
    {
     \printnames{editor}
     \setunit{\addspace}
     \printtext{editors}
     \clearname{editor}
    }
}

\renewbibmacro{in:}{}

\DeclareFieldFormat{pages}{#1}

\DeclareFieldFormat{volume}{\textbf{#1}}
\renewbibmacro*{volume+number+eid}{
    \printfield{volume}
    \iffieldundef{number}{}{\addcomma\printfield{number}}
    \setunit{\addcomma\space}
    \printfield{eid}}

\DeclareFieldFormat{doi}{\newline DOI: \href{https://doi.org/#1}{#1}}

\usepackage{amsmath}
\usepackage{graphicx}

\usepackage[T1, T2A]{fontenc}
\usepackage{setspace}
\begin{document}
\sloppy

.
\begin{center}

{\Large\textbf{Implementation of a generalized intermittency scenario in the Rössler dynamical system}}

\vspace{0.4cm}

\textbf{O.O. Horchakov, A.Yu. Shvets}

\vspace{0.4cm}

\textbf{Abstract.}
\end{center}

The realization of novel scenario involving transitions between different types of chaotic attractors is investigated for the Rössler system. Characteristic features indicative of the presence of generalized intermittency scenario in this system are identified. The properties of ``chaos–chaos'' transitions following the generalized intermittency scenario are analyzed in detail based on phase-parametric characteristics, Lyapunov characteristic exponents, phase portraits, and Poincaré sections.



\section*{Introduction}

Scenarios of generalized intermittency describe the transition from a chaotic attractor of one type to a chaotic attractor of another type. Such scenarios were initially discovered in the study of non-ideal Sommerfeld–Kononenko-type dynamical systems \cite{sommerfeld1902, kononenko1969}. These scenarios generalize the Manneville–Pomeau scenarios \cite{manneville1980a, pomeau1980}, and, in some cases, represent combinations of the Feigenbaum \cite{feigenbaum1978, feigenbaum1979} and Manneville–Pomeau scenarios. The review paper \cite{shvets2021overview} presents implementations of various versions of the generalized intermittency scenario in non-ideal pendulum, hydrodynamic, and electroelastic systems.
Moreover, transitions of the "chaos–chaos" type following the generalized intermittency scenario have also been identified in non-isolated invariant sets, the so-called maximal attractors. Strictly speaking, these sets do not qualify as attractors in the classical sense. Nevertheless, even for such atypical attracting structures, the generalized intermittency scenario can still be observed \cite{shvets2021pendulum, donetskyi2023}.

\section*{Objective and Methodology of the Study}

It was established in \cite{horchakov2024, shvets2025}, that various types of the generalized intermittency scenario can be realized in the ideal Lorenz dynamical system. The objective of the present study is to provide numerical evidence supporting the realization of the generalized intermittency scenario in such classical dynamical system as the Rössler system. The investigation employs standard techniques of chaotic dynamics, including the Runge–Kutta method for constructing phase portraits of attractors \cite{hairer1987}, the Benettin algorithm for computing the maximal Lyapunov exponent \cite{benettin1976, benettin1980}, the Hénon method for constructing Poincaré sections \cite{skiadas2016}, and a computational technique based on color-shaded encoding for visualizing the distribution of the invariant measure over the phase portrait of the attractor \cite{henon1976}. The detailed methodology for applying the above-mentioned numerical methods and algorithms is described in \cite{henon1976, kuznetsov2006, krasnopolskaya1992, shvets2012}.

\section*{Rössler System}

In~\cite{rossler1976}, a nonlinear system of three differential equations was considered:
\begin{equation}
\begin{cases}
\dot{x} = x_2 - x_3,\\
\dot{y} = x_1 + e x_2,\\
\dot{z} = f + x_3(x_1 - m),
\end{cases}
\label{eq:rossler}
\end{equation}
where $x_1, x_2, x_3$ are phase variables and $e, f, m$ are parameters of the system.

Here $x_1,x_2,x_3$ are phase variables, and e,f,m are system parameters. This system later became known as the Rössler system. It should be noted that the first two equations of system (1) are linear, while the quadratic nonlinearity appears only in the third equation. Rössler proposed this system purely heuristically, without relying on any physical assumptions in its derivation. His goal was to construct a simple deterministic third-order system of differential equations exhibiting highly complex chaotic dynamics. Over time, Rössler revisited the analysis of system (1) in his later works \cite{rossler1976, rossler1979, rossler2020}. Today, both the Rössler and Lorenz systems \cite{lorenz1963} are widely recognized as canonical examples of chaotic dynamics in low-dimensional deterministic systems.
Assume that the parameters of system (1) are $e=0.2,f=0.2$ and choose the parameter m as the bifurcation parameter. In Fig.1, a, the phase–parameter characteristic of the system is shown, constructed using the Hénon method, as the parameter m varies within the interval  $5.45<m<5.65$. Here the plane $x_2=0$  is chosen as the secant plane. Individual lines (branches) of the phase–\-param\-eter characteristic (the bifurcation tree) correspond to the limit cycles of system (1), while the densely black regions of the bifurcation tree correspond to the chaotic attractors of the system. An analysis of the phase–parameter characteristic shows that in the range $5.56 < m < 5.59$, system (1) undergoes transitions from limit cycles to chaotic attractors. These transitions occur via cascades of period-doubling bifurcations of limit cycles, that is, in full accordance with the Feigenbaum scenario \cite{feigenbaum1978, feigenbaum1979}. Once a chaotic attractor appears, it persists over a certain interval as the parameter m increases. When m reaches a certain critical value, the chaotic attractor disappears and a limit cycle again becomes the attractor of system (1). As m increases further, another transition from a regular regime to a chaotic one occurs according to the Feigenbaum scenario. It should be noted that short intervals of limit cycle existence are referred to as periodicity windows.
It should be noted that a positive maximal Lyapunov exponent is a necessary condition for the chaotic nature of a steady-state regime. Fig. 1, b shows the graph of the dependence of the maximal nonzero Lyapunov exponent $\lambda_1$ on the bifurcation parameter m. This graph was constructed using the algorithm proposed by Benettin et al. \cite{benettin1976, benettin1980}. Positive values of the Lyapunov exponent correspond to intervals of the parameter m for which chaotic attractors exist in system (1). The “drops” of the Lyapunov exponent graph into the region of negative values correspond to the periodicity windows observed in Fig. 1, a. The most interesting region of the phase–parameter characteristic (Fig. 1, a) is the neighborhood of the point $m\approx 5.585$. As seen in Fig. 1, a, in the right-side neighborhood of $m\approx5.585$, there is a significant increase in the area of the densely black region on the phase–parameter diagram. As established in \cite{shvets2021overview, shvets2025}, such an increase in the corresponding area indicates the realization of the generalized intermittency scenario. Various versions of this scenario are described in \cite{shvets2021overview, shvets2021pendulum, donetskyi2023, horchakov2024, shvets2025}. Another indication of the realization of the generalized intermittency scenario is a noticeable increase in the maximum Lyapunov exponent at $m>5.585$. We can see such increasement in Fig. 1, b.

\begin{figure}[h]
\begin{minipage}[h]{0.43\linewidth}
\center{\includegraphics[width=1\linewidth]{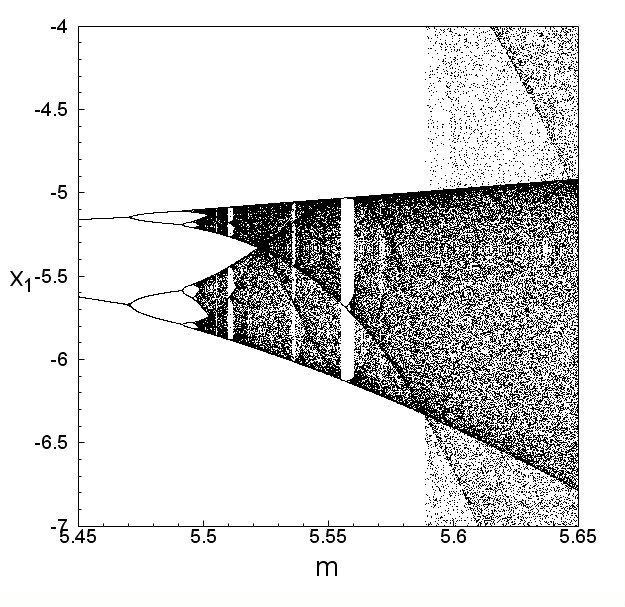}} a) \\
\end{minipage}
\hfill
\begin{minipage}[h]{0.43\linewidth}
\center{\includegraphics[width=1\linewidth]{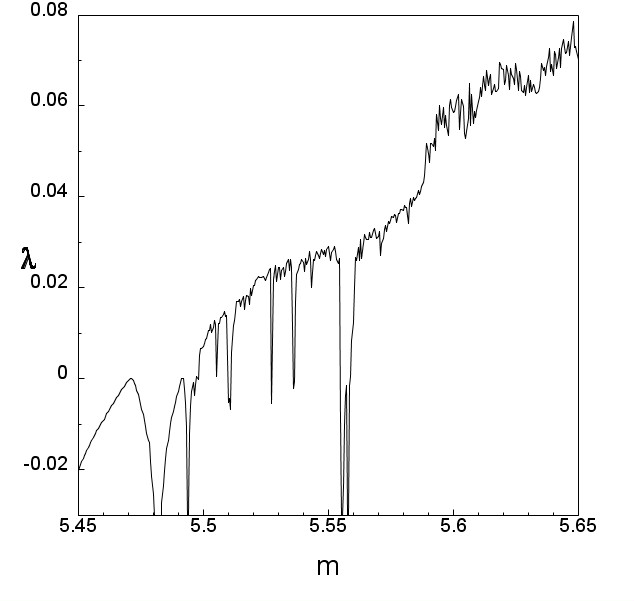}} \\b)
\end{minipage}
\vfill
\begin{minipage}[h]{0.43\linewidth}
\center{\includegraphics[width=1\linewidth]{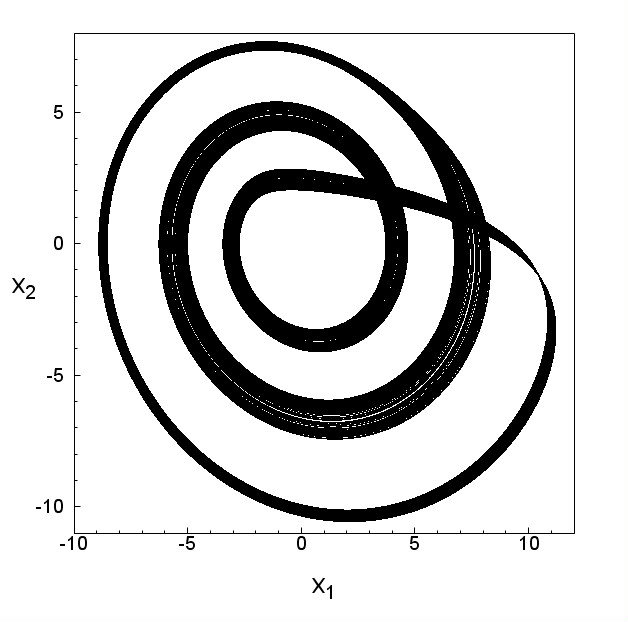} c) \\}
\end{minipage}
\hfill
\begin{minipage}[h]{0.43\linewidth}
\center{\includegraphics[width=1\linewidth]{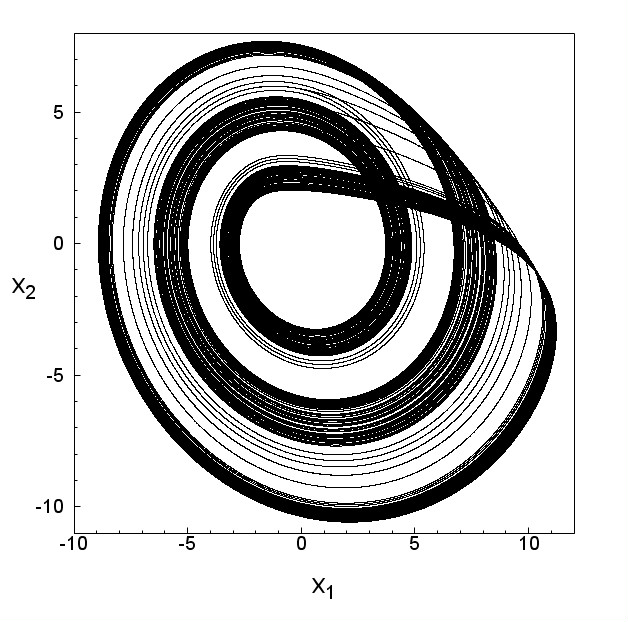}} d) \\
\end{minipage}
\caption{Phase-parametric characteristic a); maximal non-zero Lyapunov exponent b); distribution of the natural invariant measure at  $m=5.58$ (c); at $m=5.59$ (d).}
\label{ris:experimentalcorrelationsignals}
\end{figure}

Let us now examine in more detail the realization of the generalized intermittency scenario in the Rössler system by analyzing the distributions of natural invariant measures and Poincaré sections.

In Fig. 1, c is shown the distribution of the invariant measure over the phase portrait of the chaotic attractor at $m=5.58$. As the parameter m increases, a hard bifurcation occurs in system (1), as a result of which the existing chaotic attractor disappears and a new type of chaotic attractor emerges. The distribution of the invariant measure over the phase portrait of this new chaotic attractor, constructed at $m=5.59$, is shown in Fig. 1, d. The distributions of the invariant measure were constructed using the algorithm of computer encoding in shades of black \cite{henon1976, kuznetsov2006}. The trajectory motion along the new chaotic attractor exhibits phase alternation between two phase - a coarse-grain (rough) laminar phase and a turbulent phase. The coarse-grain laminar phase corresponds to chaotic wanderings of the trajectory in the region of localization of the disappeared chaotic attractor (dense black region in Fig. 1, c). At an unpredictable moment in time, the trajectory leaves the localization region of the vanished chaotic attractor and “escapes” to more distant areas of the phase space (gray points in Fig. 1, d). Such motions correspond to the turbulent phase of intermittency. Alternations between the coarse-grained laminar phase and the turbulent phase are observed an infinite number times. The transition time from one phase to another is also unpredictable. On average, the duration of the coarse-grained laminar phase exceeds that of the turbulent phase. This process fully corresponds to the scenario of generalized intermittency \cite{shvets2021overview, horchakov2024, shvets2025}. The scenario of generalized intermittency can also be identified by analyzing the Poincaré sections. In Fig. 2, the Poincaré sections of chaotic attractors at $m=5.58$ and $m=5.59$ are constructed using the Hénon method. Both sections exhibit a quasi-ribbon structure and represent chaotic sets of discrete points. It is worth noting that such a quasi-ribbon structure is characteristic of chaotic attractors in the Rössler system. As shown in Fig. 2, b, the structure of the Poincaré section at $m=5.59$ contains all the fragments present in the Poincaré section of the chaotic attractor at $m=5.58$  (Fig. 2, a).

\begin{figure}[h]
\begin{minipage}[h]{0.5\linewidth}
\center{\includegraphics[width=1\linewidth]{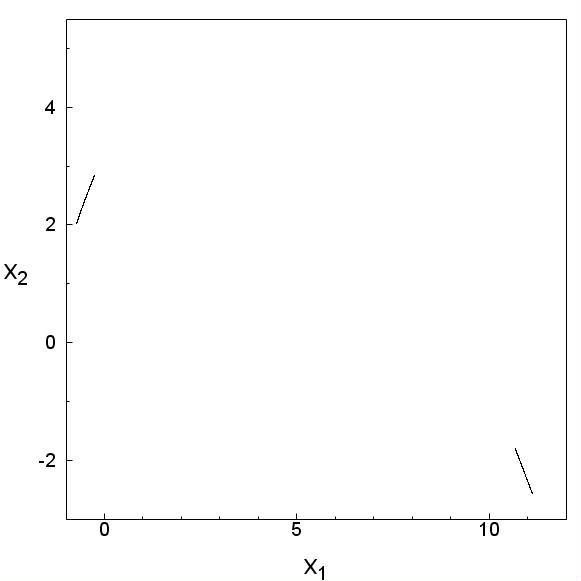} \\ а)}
\end{minipage}
\hfill
\begin{minipage}[h]{0.5\linewidth}
\center{\includegraphics[width=1\linewidth]{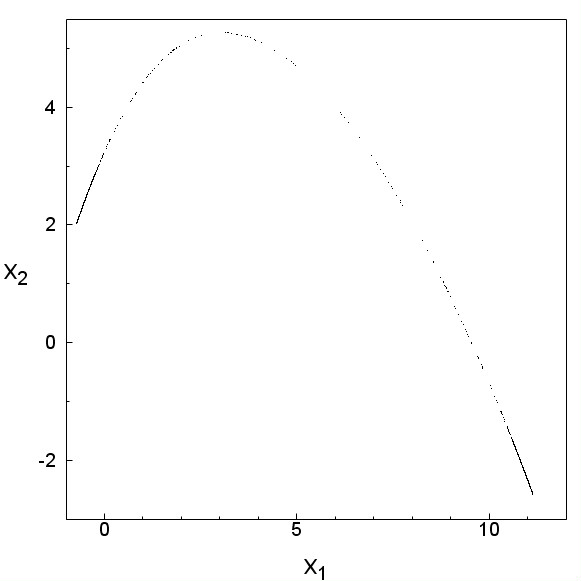} \\ b)}
\end{minipage}
\caption{Poincaré sections at $m=5.58$ (a), at $m=5.59$ (b)}
\label{ris:image1}
\end{figure}

These fragments form the coarse-grained laminar phase of the attractor at~$m=5.59$. Accordingly, new points appear in the Poincaré section at $m=5.59$, corresponding to the turbulent phase.  
Let us now consider the bifurcations in the Rössler system as the parameter m varies within the interval $(16,18.5)$. The values of the parameters e and f remain unchanged. As before, using the methods of Hénon, Benettin, and computer-based color coding, we construct a series of dynamic characteristics of the Rössler system. Thus, in Fig. 3, a, the phase–parameter characteristic of the Rössler system is presented. The constructed bifurcation tree provides a clear representation of the types of attractors in system (1). The individual branches of the bifurcation tree correspond to limit cycles, while the densely black regions of the tree represent chaotic attractors. Moreover, this figure makes it possible to identify transition scenarios, including both “limit cycle-to-chaos” and “chaos-to-chaos” transitions.
The constructed bifurcation tree demonstrates a symmetry in the transitions to chaos, both with increasing and decreasing values of the parameter m. As m  increases, starting from $m=16.7$, in the system begins an infinite cascade of period-doubling bifurcations of limit cycles, followed by the emergence of a chaotic attractor with a relatively small localization region in the phase space. This represents a transition to chaos following the Feigenbaum scenario. A similar scenario is observed as m decreases, beginning from $m=18.05$. Particular attention should also be paid to two bifurcation points: $m \approx17.35$ and $m\approx17.795$.

\begin{figure}[h]
\begin{minipage}[h]{0.43\linewidth}
\center{\includegraphics[width=1\linewidth]{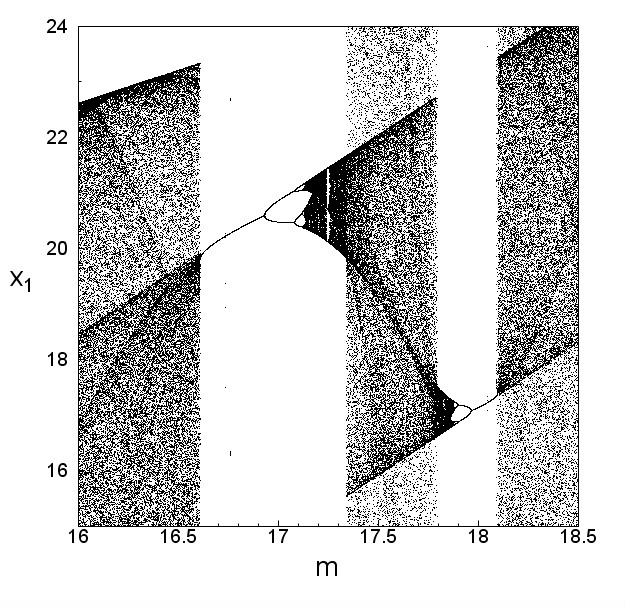}} a) \\
\end{minipage}
\hfill
\begin{minipage}[h]{0.43\linewidth}
\center{\includegraphics[width=1\linewidth]{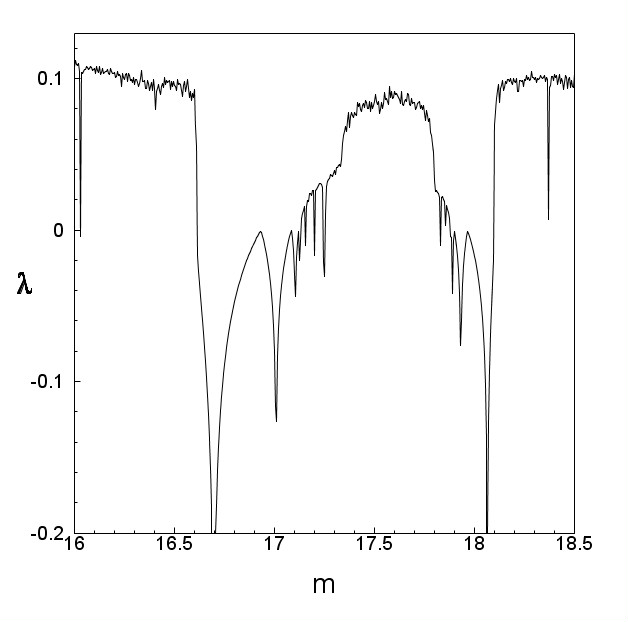}} \\b)
\end{minipage}
\vfill
\begin{minipage}[h]{0.43\linewidth}
\center{\includegraphics[width=1\linewidth]{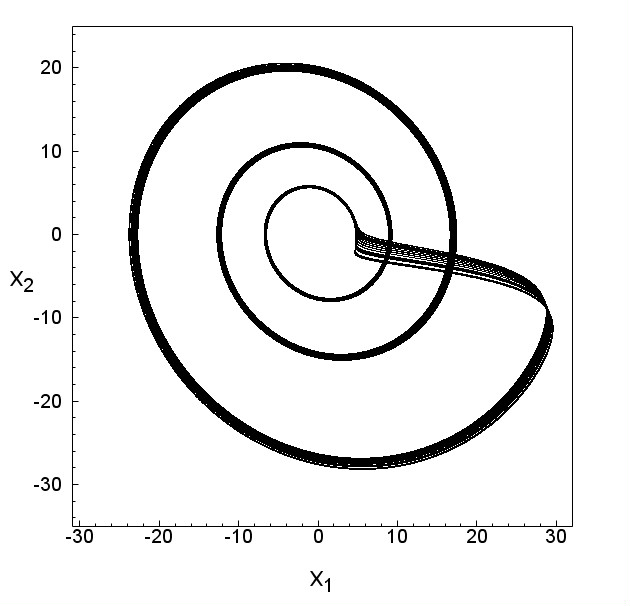} c) \\}
\end{minipage}
\hfill
\begin{minipage}[h]{0.43\linewidth}
\center{\includegraphics[width=1\linewidth]{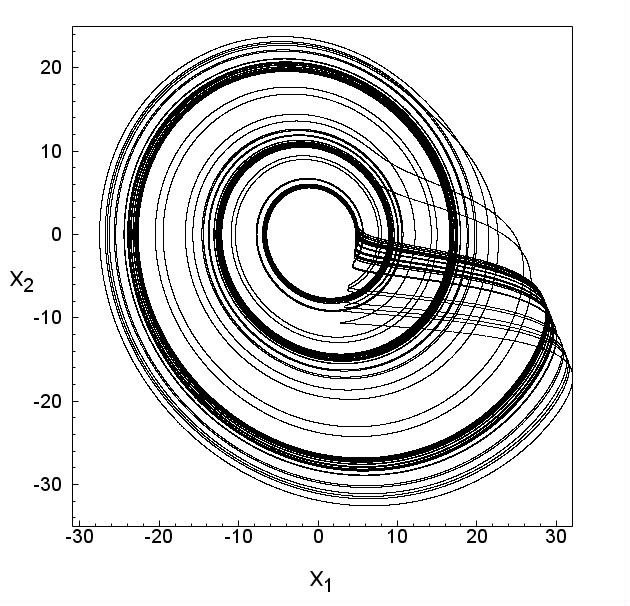}} d) \\
\end{minipage}
\caption{Phase-parametric characteristic (a), Maximal non-zero Lyapunov exponent (b) and projections of distribution of the natural invariant measure at $m=17.8$ (c), at $m=17.79$ (d)}
\label{ris:experimentalcorrelationsignals}
\end{figure}

In the right-hand neighborhood of $m\approx17.35$ (and the left-hand neighborhood of $m\approx17.795$), a significant increase in the area of the densely black chaotic region in Fig. 3, a is observed, indicating the realization of a generalized intermittency scenario of the transition from one type of chaotic attractor to another. 
In addition, two bifurcation points are clearly visible at $m\approx16.6$ and $m\approx18.1$. As the system passes through these points, a “limit cycle–chaos” transition occurs following the Pomeau–Manneville scenario. In Fig. 3, b, the graph of the dependence of the maximal nonzero Lyapunov exponent on the bifurcation parameter m is presented. As seen from the graph, for $m>17.35$ and $m<17.795$, the value of the maximal Lyapunov exponent nearly doubles. This increase is further evidence of the realization of a generalized intermittency scenario in the Rössler system.
Finally, let us consider the realization of the generalized intermittency scenario through the phase portraits of chaotic attractors of different types. 
In Fig. 3, c, the projection of the invariant measure distribution for the chaotic attractor at $m=17.8$ is shown, while Fig. 3, d presents the projection of the invariant measure distribution for the chaotic attractor at $m=17.79$. As the value of the parameter m decreases, the chaotic attractor that existed in the right-hand neighborhood of the bifurcation point $m=17.795$ disappears, and for $m<17.795$, a new type of chaotic attractor emerges. The motion of trajectories on this new attractor includes two phases, clearly identifiable in Fig. 3, d: a coarse-grain laminar phase and a turbulent phase. In the coarse-grain laminar phase (the densely black fragment in Fig. 3, d), the trajectory performs chaotic wandering in a neighborhood of the phase-space localization region of the attractor that existed for  $m>17.795$. The turbulent phase (the gray fragments in Fig. 3, d) corresponds to the trajectory’s excursions into more distant regions of the phase space.
Similarly, the generalized intermittency scenario can be illustrated through Poincaré sections, as was done in Fig. 2. It should be noted that, in contrast to the previously analyzed case, the transition to chaos through the generalized intermittency scenario can occur both with increasing and decreasing values of the parameter m.
The implementation of the generalized intermittency scenario can also be observed in other regions of the parameter space of the Rössler system. Let us assume that $e=0.2$ and $m=17.4$, while the bifurcation parameter is chosen to be $f$. 

\begin{figure}[h]
\begin{minipage}[h]{0.43\linewidth}
\center{\includegraphics[width=1\linewidth]{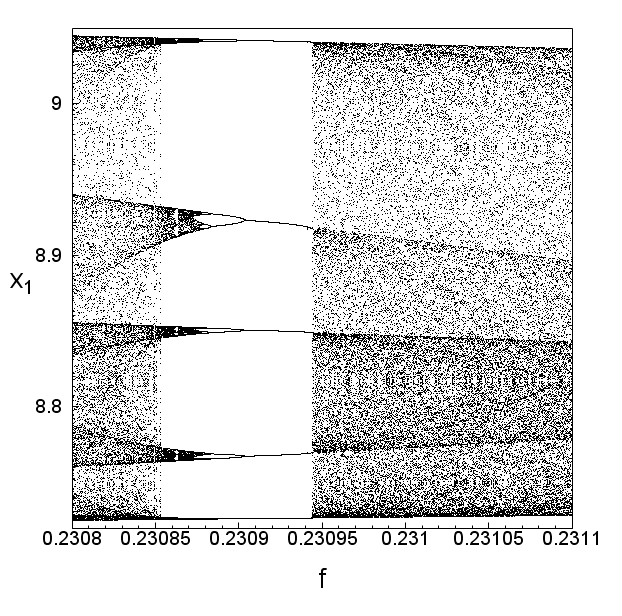}} a) \\
\end{minipage}
\hfill
\begin{minipage}[h]{0.43\linewidth}
\center{\includegraphics[width=1\linewidth]{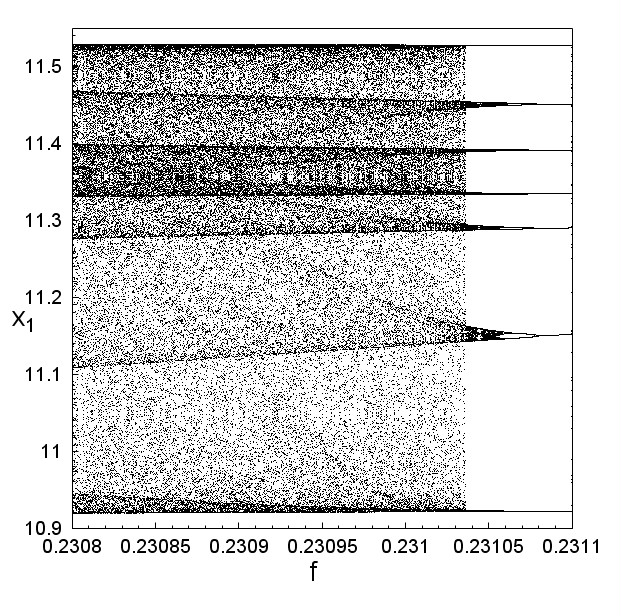}} \\b)
\end{minipage}
\vfill
\begin{minipage}[h]{0.43\linewidth}
\center{\includegraphics[width=1\linewidth]{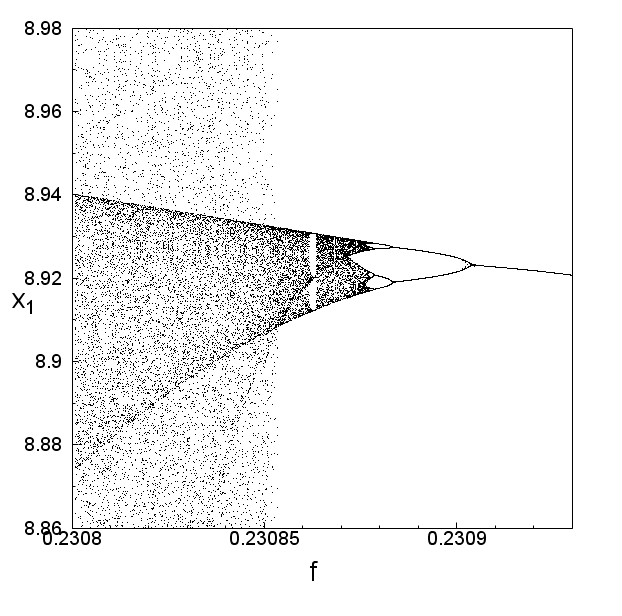} c) \\}
\end{minipage}
\hfill
\begin{minipage}[h]{0.43\linewidth}
\center{\includegraphics[width=1\linewidth]{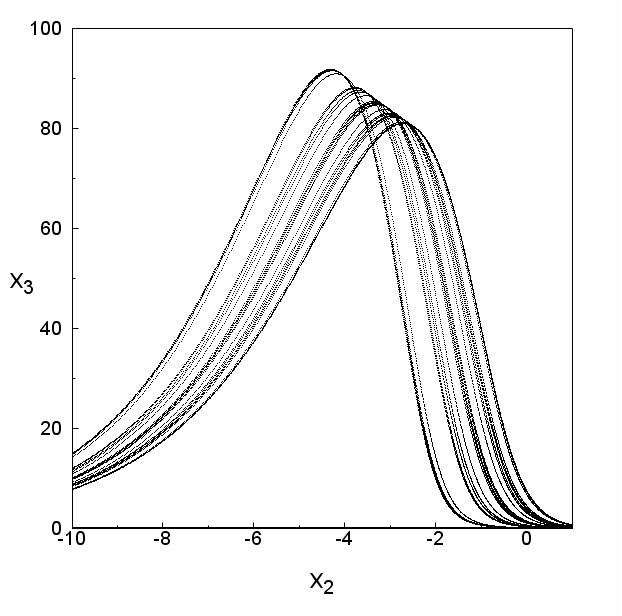}} d) \\
\end{minipage}
\caption{Phase-parametric characteristics(a-c) and fragment of distribution of invariant measure at $f=0.23082$  (d)}
\label{ris:experimentalcorrelationsignals}
\end{figure}

We will investigate the dynamical behavior of system (1) within the range\\ $0.2308<f<0.2311$. For these parameter values, the Rössler system has two coexisting attractors, each possessing its own basin of attraction. Fig. 4, a,b presents the phase–parameter characteristics of two different attractors constructed using the Hénon method.
As before, individual branches of the bifurcation trees correspond to limit cycles, while the densely black regions represent chaotic attractors. Despite a certain similarity between these phase–parameter characteristics, it is clearly seen—by examining the intervals of variation of the coordinate $x_1$—that the corresponding attractors are localized in different regions of the phase space.
Let us now focus exclusively on the realization of the generalized intermittency scenario. As noted earlier, an indicator of this scenario is a significant increase in the area of the densely black (chaotic) regions on the phase–parameter characteristic. Such increases in the areas of the densely black regions can be observed on both phase–parameter characteristics. This indicates the possibility of a transition “chaotic attractor of one kind → chaotic attractor of another kind” according to the generalized intermittency scenario.
Let us examine this scenario in more detail using one of the coexisting attractors as an example. Fig.4, c shows a fragment of the phase–parameter characteristic from Fig.4, a. The enlarged scale in Fig.4, c makes it possible to identify the bifurcation point $f=0.23085$, at which a “chaos → chaos” transition occurs according to the generalized intermittency scenario. In Fig.4, d is shown an enlarged fragment of the distribution of the invariant measure over the phase portrait of the attractor at $f=0.23082$. This attractor appears as the parameter f decreases immediately after the bifurcation point $f\approx0.23085$. The use of the enlarged scale makes it possible to clearly visualize the features of this distribution. One can distinguish a coarse-grain laminar phase of the trajectory (the densely black region in the figure) and a turbulent phase (the gray-shaded areas).
Let us emphasize once again that the coarse-grain laminar phase almost coincides with the region of localization in phase space of the chaotic attractor that exists for $f>0.23085$ and disappears after the bifurcation point is passed. Another confirmation of the generalized intermittency scenario is a noticeable increase in the value of the maximal Lyapunov exponent. Specifically, for the chaotic attractor at $f=0.23086$, the maximal Lyapunov exponent is $\lambda_1=0.005$, while for the chaotic attractor at $f=0.23082$, it increases to $\lambda_1=0.010$.

\section*{Conclusions}

Thus, the generalized intermittency scenario, previously identified in non-ideal dynamical systems, is also realized in ideal dynamical systems such as the classical ideal Rössler system. Future research will focus on identifying the realization of other types of the generalized intermittency scenario in various ideal dynamical systems.

\section*{Acknowledgments}

This work was supported by a grant from the Simons Foundation International (SFI-PD-Ukraine-00014586, O.O. Horchakov).

\printbibliography

\textit{O.O. Horchakov}, Institute of Mathematics of the National Academy of Sciences of Ukraine, st. Tereschenkivska 3, 01024 Kyiv, Ukraine 
o.horchakov@imath.kiev.ua

\textit{A.Yu. Shvets}, Institute of Mathematics of the National Academy of Sciences of Ukraine, st. Tereschenkivska 3, 01024 Kyiv, Ukraine oshvets@imath.kiev.ua

\end{document}